\documentclass[letter,referee]{aa}  
\usepackage{graphicx}
\usepackage{txfonts}
\usepackage[colorlinks=true,citecolor=blue,breaklinks=true]{
hyperref}
\bibpunct{(}{)}{;}{a}{}{,} 
\usepackage[switch,pagewise]{lineno}
\usepackage{xspace}
\newcommand{\alfven}{Alfv\'en\xspace}
\newcommand{\unit}[1]{\ensuremath{\,\mathrm {#1}}}  
\newcommand{\figref}[1]{Figure~\ref{#1}}
\newcommand{\secref}[1]{Section~\ref{#1}}
\renewcommand{\eqref}[1]{Equation~\ref{#1}}

\renewcommand{\deg}{\ensuremath{^\circ}}

\begin{document}

   \title{Abnormal oscillation modes in a waning light bridge}

   \author{Ding Yuan\and Robert W. Walsh}
   \institute{Jeremiah Horrocks Institute, University of Central Lancashire,
    Preston PR1 2HE, UK \\ \email{DYuan2@uclan.ac.uk}}
   \date{}

 
  \abstract
   {A sunspot acts as a waveguide in response to the dynamics of the solar interior; the trapped waves and oscillations could reveal its thermal and magnetic structures.}
   {We study the oscillations in a sunspot intruded by a light bridge, the details of the oscillations could reveal the fine structure of the magnetic topology.}
   {We use the Solar Dynamics Observatory/Atmospheric Imaging Assembly data to analyse the oscillations in the emission intensity of light bridge plasma at different temperatures and investigate their spatial distributions.}
   {The extreme ultraviolet emission intensity exhibits two persistent oscillations at five-minute and sub-minute ranges. The spatial distribution of the five-minute oscillation follows the spine of the bridge; whereas the sub-minute oscillations overlap with two flanks of the bridge. Moreover, the sub-minute oscillations are highly correlated in spatial domain, however, the oscillations at the eastern and western flanks are asymmetric with regard to the lag time. In the meanwhile, jet-like activities are only found at the eastern flank.
}
   {\textbf{Asymmetries} in forms of oscillatory pattern and jet-like activities \textbf{are} found between two flanks of a granular light bridge. Based on our study and recent findings, we propose a new model of twisted magnetic field for a light bridge and its dynamic interactions with the magnetic field of a sunspot.}


   \maketitle
%

\section{Introduction}

A sunspot is a localized dark and cool area on the Sun with strong magnetic field that suppresses the dynamic motions of the hot gas. Waves and oscillations are established within a sunspot, which act as a waveguide in response to the dynamics and turbulence of the solar interior \citep[see a comprehensive review by][]{khomenko2015}. Analysis of the spatial and height distributions of the oscillations can reveal a sunspot's magnetic \citep{reznikova2012,yuan2014sp,yuan2014lb,jess2013,jess2016} and thermal structures \citep{zhugzhda2008,botha2011,snow2015,yuan2016st,chae2015,kwak2016}. An isolated sunspot could be assumed to have an expanding axisymmetric magnetic field; inhomogeneities, such as light bridges, umbral dots, are normally observed at a variety of scales \citep[see a review by ][]{borrero2011}. There inhomogeneities could modify the distribution of the oscillation frequency \citep{jess2012} and even even cause abnormal frequency changes  \citep{yuan2014lb,su2016}. Therefore, they are excellent objects to test the feasibility and robustness of the presumed theory and to explore the induced extra dynamics, e.g., reconnections, heating, flows, and convections.

This study concentrates on one such inhomogeneities: light bridge.
A light bridge is usually formed as a lower atmospheric structure in nascent or decaying sunspots. It divides an umbra into separate cores \citep{sobotka1994,lagg2014}, and convection, which is normally suppressed by a sunspot's strong magnetic field, is partially restored \citep{rimmele199,schussler2006}. Consequently, upflows are usually observed at the spine of a bridge, and downflows (or return flows) at the two flanks \citep{louis2009}. The downflows drag the umbral field downward and may \textbf{even} cause polarity reversal \citep{lagg2014}. Recent numerical simulations reveal that light bridges could be formed due to an emerging bipole field \citep{toriumi2015b,toriumi2015a}. Magnetic null points formed at a flank of a light bridge could be preferred locations for magnetic reconnection \citep{cheung2012}. \citet{louis2014} found that one flank of a bridge supports prevailing chromospheric jet activities caused mainly by magnetic reconnection. This implies the magnetic field of a light bridge may be twisted along its length and the field may only structure a magnetic null point at one flank.

\citet{yuan2014lb} reported the first detection of five minute oscillation at a light bridge observed at the chromospheric height. \citet{su2016} suggested that the inference counterstream three minute umbral wave might be the source of the five-minute oscillations at light bridge. \citet{yang2015} detected transverse oscillation of an wing of a light bridge at five minute range and termed it as ``light wall oscillation''. This kind of oscillation could be detected as periodic intensity variations if the motion is along the line-of-sight \citep{yuan2016sv,yuan2016as}. At current stage, the source of the five minute oscillations at light bridges still awaits to be revealed.

In this letter, we present the first detection of two abnormal oscillation modes in a waning light bridge. These  modes are anomalous in comparison \textbf{with the} conventional sunspot oscillations \citep{jess2016,khomenko2015,yuan2014sp}, and are uniquely  associated with granular light bridges. The spatial distributions of the oscillations reveal the fine structure of the bridge, potential heating mechanism that sustains a significant portion of the plasma at coronal temperature, and a possible magnetic ablation process that eventually leads to the disappearance of the bridge. We present the method in \secref{sec:method} and results and discussions in \secref{sec:results}.

\section{Method}
\label{sec:method}
\subsection{Data reduction}
This study focuses on active region AR 11520 located at the southern hemisphere on 11 July 2012. The main spot has a positive polarity and is formed of two umbral cores separated by a light bridge (\figref{fig:fov}). The bridge under examination is a photospheric (or granular) light bridge \citep{muller1979,sobotka1994}; it has a central lane with relatively stronger magnetic field component ($B_z$) along the local normal direction than either flank (\figref{fig:fov}(e) and \figref{fig:prof}). A central lane is usually dark if observed in G-band \citep{rimmele2008}. Upflows are detected at the spine of a bridge and surrounded by weaker downflows at two flanks \citep{rouppevandervoort2010}. Three days later, two umbral cores eventually coalesced into a single umbra (\figref{fig:fov}(e-g)). During this process, the bridge waned and eventually vanished. 

It is unusual that a light bridge, which is normally a low-lying feature at photospheric and chromospheric heights, is also observed at coronal temperatures \citep{mathews2012}. In this case, the light bridge is clearly visible in the 131 \AA{}, 171 \AA{}, 193 \AA{}, 211 \AA{} channels, and marginally seen in the 94 \AA{} and 335 \AA{} bandpasses, so it is very unlikely that the cool components of the response functions made significant contributions to the coronal line emissions \citep{boerner2012}. Therefore, we believe a significant portion of the bridge's plasma is heated to coronal temperatures. The bridge's coronal component is clear of background contamination, because the sunspot has an overwhelming positive polarity, and that the the majority of field lines extend radially, rather than connect to nearby negative polarities to form coronal loops (\figref{fig:fov}). Consequently, this could be considered as a unique experiment to investigate magnetic field interactions at an intermediate spatial scale.

One hour observation was examined starting at 18:30:00 UT 11 July 2012; the data are provided by the Atmospheric Imaging Assembly \citep[AIA,][]{lemen2012} onboard the Solar Dynamics Observatory \citep[SDO,][]{pesnell2012}. The analyse concentrates upon the $1600\unit{\AA{}}$, $171\unit{\AA{}}$ and $211\unit{\AA{}}$ channels, which have nominal response temperatures of $5,000\unit{K}$, $600,000\unit{K}$, and $2,000,000\unit{K}$, respectively. The extreme ultraviolet (EUV, $171\unit{\AA{}}$ and $211\unit{\AA{}}$) channels both take an image every 12 seconds; while the UV ($1600\unit{\AA{}}$) channel capture a frame every 24 seconds. AIA level 1 data were obtained and has already been processed by removing the bad pixels and spikes and correcting the flat field. We used the standard routine (prep$\_$aia.pro) in the SolarSoft\footnote{\url{http://www.lmsal.com/solarsoft/}} to calibrate the data into level $1.5$. During this step, the data were corrected for the roll angles, interpolated into a common plate scale of $0.6\arcsec$, and normalized by the exposure time. A small field-of-view (FOV, \figref{fig:fov}(b-d)) was traced against solar differential rotation and was co-aligned to a sub-pixel accuracy. 

The magnetograms ($B_z$) (\figref{fig:fov}(e-g)) are provided by the Stokes Spectro-Polarimeter (SP), a component of the Solar Optical Telescope (SOT) \citep{tsuneta2008} aboard the Hinode satellite \citep{kosugi2007}. SOT/SP provides Full-Stokes measures of two \ion{Fe}{1} lines at $630.15\unit{\AA{}}$ and $630.25\unit{\AA{}}$ at the photosphere. Three raster data sets of SOT/SP are used (\figref{fig:fov}). The raster scans started at 17:35:28 UT 11 July 2012, 00:47:08 UT 13 July 2012, and 16:00:42 UT 14 July 2012, respectively; and each scan lasted for about 24 minutes. The raster scans have spatial sampling of about $0.3\arcsec$ and step size of $0.32\arcsec$. We used the SOT/SP level 2 data that are the output of full Milne-Eddington inversion\footnote{The level 2 MERLIN inversions are available at \url{https://www2.hao.ucar.edu/csac/csac-data}.} \citep{lites2007}. The $180\deg$ ambiguity in the azimuthal angle of the magnetic field line is corrected by using the AZAM package \citep{lites1995}. 

We use a Cartesian reference frame $[x,y,z]$ with $\vec{z}$ along the local normal direction. The vertical electric current gives a quantitative measure to the disruption of horizontal magnetic field and is calculated as $j_z=(\partial B_y/\partial x - \partial B_x/\partial y)/\mu_0$ (see the contour in \figref{fig:pm}(a)), where $\mu_0$ is the magnetic permeability in free space.

Hinode/SOT Broadband Filter Imager (BFI) provides \ion{Ca}{II} H-line observations that are sensitive to emission of plasmas at chromospheric temperatures. Unfortunately, BFI only take images of sunspot AR 11520 at a best cadence of 5 minutes. Twenty-three images were taken between 16:00 - 18:10 UT on 11 \textbf{Jul} 2012, within which six jet-like events were spotted by identifying transient elongated features protruding from the bridge \citep[see method in][]{louis2014}. Another five events were detected within twenty-seven images captured between 20:20-24:00 UT. The bases of the jet-like activities are marked in \figref{fig:pm}(d).

\subsection{Time series and spectral analysis}

To illustrate the presence of oscillations and their spectrum dynamics, the emission intensity variations at the western flank of the light bridge were extracted, these are represented by the co-spatial pixels in the $1600\unit{\AA{}}$, $171\unit{\AA{}}$ and $211\unit{\AA{}}$ images (the crosses in \figref{fig:fov}(b-d)). The long period trend in each time series was removed by subtracting a moving average of five minute interval. Then the de-trended time series were normalized by its standard deviation. In \figref{fig:ts}(a), they are plotted subsequently by offsetting their mean values arbitrarily to 2, 6, and 10, respectively, for visualization purposes. The wavelet spectra (\figref{fig:ts}(b-d)) were calculated using the Morlet mother function, which is optimal for illustrating oscillatory processes \citep{torrence1998}. 

To investigate the spatial distribution of the oscillatory processes, a power map was constructed for each significant spectral component in a smaller region as enclosed in the rectangle of \figref{fig:fov}(b). Times series of each pixel was de-trended as in the same manner as for the reference pixel analysed in \figref{fig:ts}. The Fourier spectrum was computed using Fast Fourier Transform. The power was averaged over two bands at $0.5-1\unit{minute}$ and $3-5\unit{minute}$, respectively (see \figref{fig:pm}(a) and (d)). 

To investigate the coherence between different locations, we calculate the cross-correlation coefficient (XCC, \figref{fig:pm}(b)) and lag time (\figref{fig:pm}(c)) between the time series of each pixel and that of the reference pixel (labeled by a cross in \figref{fig:fov}(b)). Only the pixels with oscillation power above the 10\% level relative to the maximum power of the power map of $0.5-1\unit{minute}$ are considered. The XCCs have a minimum value of about 0.4, meaning the oscillations are highly correlated within the light bridge.

To study the coherence and lag time between the oscillations at five minute range, we use the five-point (one minute) running average of the time series, and then re-calculate the XCC and lag time (\figref{fig:pm}(e) and (f)). The running averaging removes  the sub-minute oscillations, therefore, the XCC and lag time measured in this step reflect the coherence and sequential order of the five minute oscillations.

\section{Results and Discussions}
\label{sec:results}
\begin{figure}[ht!]
\centering
\includegraphics[width=0.48\textwidth]{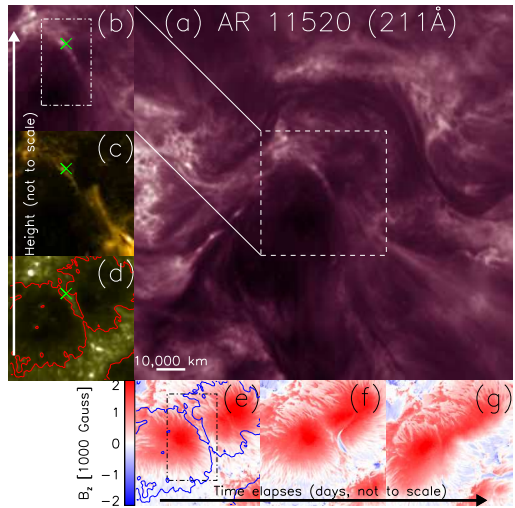}
\caption{(a) Field-of-view (FOV) of active region AR 11520 observed in the AIA 211 \AA{} channel on 11 July 2012. The active region is relatively clear of coronal loops and other complex coronal structures, therefore, the light bridge is easily identifiable. Smaller FOV (as enclosed in the dashed rectangle in (a)) observed in the AIA (b) $211\unit{\AA{}}$ (2,000,000 K), (c) $171\unit{\AA{}}$ (600,000K), and (d) $1600\unit{\AA{}}$ (5000 K) channels, respectively. The green crosses label the pixel where time series were extracted and analysed in \figref{fig:ts}. The dash-dot rectangle in (b) mark the area where power maps are calculated and analysed in \figref{fig:pm}.
(e)-(g) Evolution of the local normal magnetic field measured by Hinode SOT/SP on 11, 13, and 14 July 2012, respectively. The contours in (d) and (e) show the light bridge structure at upper photosphere ($1600\unit{\AA{}}$) as a reference. \label{fig:fov}}
\end{figure}

\begin{figure}[ht]
\centering
\includegraphics[width=0.4\textwidth]{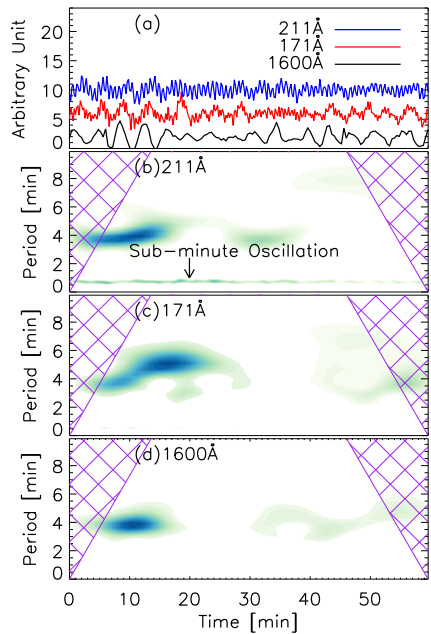}
\caption{(a) De-trended and normalized emission intensity variations at three AIA bandpasses at the location labeled in \figref{fig:fov}. (b) - (d) Wavelet power spectrum for the  $211\unit{\AA{}}$, $171\unit{\AA{}}$, and $1600\unit{\AA{}}$ bandpasses, respectively. Only the power above 90\% confidence level are illustrated. The cross-hatched regions are unreliable owing to zero padding used in the wavelet transform. \label{fig:ts}}
\end{figure}

Light curves of the emission intensity variations are extracted at a representative location of  the western flank of the light bridge (\figref{fig:fov}). Five-minute oscillations ($3-5$ minutes) are significant in all bandpasses considered (\figref{fig:ts}); this phenomenon was reported earlier \textbf{at chromospheric heights} by \citet{yuan2014lb} and \citet{su2016}. Another significant spectral component is detected at $0.5-1$ minute (sub-minute oscillations), but only in the $211$ \AA{} channel\footnote{We also detect significant sub-minute oscillations in the 193 \AA{} channel, since it is redundant compared to the 211 \AA{} data, we decided not to include it.} (\figref{fig:ts}(b)). It implies that only the plasmas with temperature between $600,000\unit{K}$ and $2,500,000\unit{K}$ are involved. The lack of the sub-minute oscillation in the 1600 \AA{} bandpass could owe to the fact that the 1600 \AA{} channel cannot detect period shorter than 48 s. However, we did not detect sub-minute oscillations either in the 304 bandpass, which has a lower detection limit of 24 s. Therefore, we are convinced that the sub-minute oscillations \textbf{do} not appear in the cool channels.

\begin{figure}[ht]
\centering
\includegraphics[width=0.45\textwidth]{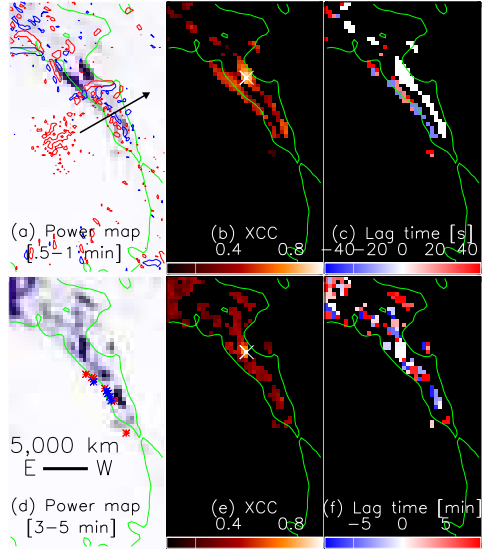}
\caption{(a) and (d) Distributions of the Fourier powers of the sub-minute and five-minute oscillations. (b) and (c) Distributions of XCC and lag time of the sub-minute oscillations; (e) and (f) are the counterparts of the five minute oscillations.
The green contour in each panel provides a reference of the light bridge structure at the upper photosphere. The red and blue contours in (a) labels the electric current density $j_z=\pm40\unit{mA\cdot m^{-2}}$, respectively. The red and blue asterisks in (d) denotes the base of the jet activities detected at time intervals of 16:00 - 18:10 UT and 20:20 - 24:00 UT, respectively.
\label{fig:pm}}
\end{figure}

\figref{fig:pm}(a) and (d) shows the spatial distributions of the sub-minute and five-minute oscillations measured in the $211$ \AA{} channel. The five-minute oscillations are distributed along the spine of the light bridge; whereas the sub-minute oscillations are distinctively co-spatial with the two flanks of the bridge. It strongly indicates that the driver for these sub-minute oscillations is likely to be connected with the collective interactions between magnetic field of the bridge and the sunspot.

\figref{fig:pm}(b) shows that the sub-minute oscillations are highly correlated with each other; the XCC is generally greater than 0.4. \figref{fig:pm}(c) illustrates that the sub-minute oscillations along the western flank of the bridge have zero-time lag. Therefore, it could be postulated that these emission intensity variations are driven by surface-to-surface buffering, rather than random interactions of adjacent fieldlines along \textbf{that} flank. The emission intensity variations at the eastern flank respond either 24 seconds ahead or after the signal of the western flank. Two pairs of surface-to-surface interactions may exist at the eastern flank, however, the AIA resolution does not allow a conclusive distinction.

Therefore, it can be inferred that the sub-minute oscillation at two ridges could have the same driver with regard to the periodicity, as they are highly correlated. However, the processes occurring at two flanks might be different, based on the fact that the lag time distribution shows asymmetry in spatial domain. It should be noted that there is some weak evidence that the eastern flank is surrounded by isolated islands of strong electric current, while the western flank is relatively much clear of such islands (\figref{fig:pm}(a)). Moreover, the jet-like activities are only detected at the eastern flank\footnote{Although SOT/BFI observations did not cover the AIA observation interval used in this study, we don't think the spatial distribution of the jet activities will change within a timescale of hours.} (\figref{fig:pm}(d)). It means magnetic reconnection is repetitive triggered there.

\begin{figure}[ht]
\centering
\includegraphics[width=0.4\textwidth]{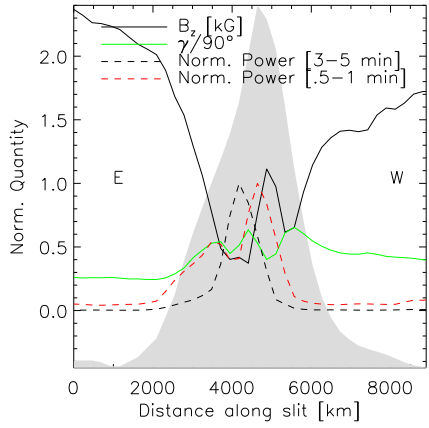}
\caption{Profiles of local normal magnetic field strength $B_z$, field inclination $\gamma/90\deg$, and the normalized powers of five and sub-minute oscillations along the arrow labeled in \figref{fig:pm}(a). The grey shade plots the intensity variation of the 1600 \AA{} (scaled to the plot range), and provides a reference for the position of the light bridge.
\label{fig:prof}}
\end{figure}

\figref{fig:prof} plots the profiles of local normal magnetic field strength $B_z$, the field inclination $\gamma$, and the normalized powers of the sub- and five minute oscillations across the bridge, as labeled by an arrow in \figref{fig:pm}(a). Both profiles of $B_z$ and $\gamma$ have two local valleys across the bridge. At the western flank, the decrease (increase) of $B_z$ is associated with the increase (decrease) of $\gamma$. This anti-correlation is expected if we consider that $B_z=B\cos(\gamma)$ for $\gamma\in[0,90\deg]$. In contrast, at the eastern flank, such anti-correlation does not hold, meaning that more complex magnetic structure may have formed and that the field strength and azimuthal angle also have significant influence on the structure. It supports the asymmetries found in the spatial pattern of sub-minute oscillations and jet activities. However, at current stage, we cannot reach better conclusion with limited information.

\figref{fig:pm}(e) and (f) shows the XCC and lag time of the five-minute oscillations. In contrast, the XCC is relative weaker, and lag time forms localized patches. It means the coherence length is much shorter than that of sub-minute oscillations. \citet{yuan2014lb} found that five minute oscillations are almost in phase at chromosphere, so it is still unknown why the five-minute oscillations lose spatial coherence when observed in coronal lines.

\begin{figure}[ht]
\centering
\includegraphics[width=0.45\textwidth]{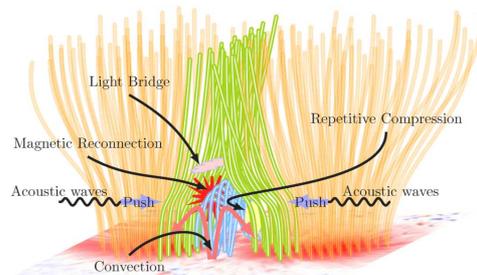}
\caption{Sketch illustrating the magnetic morphology and dynamics. The magnetic field (sky blue) of the light bridge is weakly twisted and intrudes into the sunspot. The field lines (dark green) sandwiching the light bridge are dragged by the downward motions of partial convections at two flanks of the bridge and become more inclined. Compression and rarefaction dominate the western flank, while magnetic reconnection is triggered at the eastern flank. 
\label{fig:sp}}
\end{figure}

Recent simulations and observations of \citet{toriumi2015b,toriumi2015a} show that the magnetic field of a light bridge is twisted and related to an emerging dipole field \citep[also supported by][]{louis2015}. Based on this knowledge and our observation, we propose a scenario to explain these results as depicted in \figref{fig:sp}. The sunspot has a predominant radially expanding magnetic field; the magnetic field of the light bridge is slightly twisted and intrudes into the umbral field. The buffering between two magnetic fields are driven by any dynamics of the ambient fluid: convections, granulation, turbulence, seismic waves, jet activities, self-organized dynamics, etc. At the western flank, the bridge's magnetic field does not reverse direction relative to the umbral field, therefore, compression and rarefaction between two fields dominate and cause temperature perturbations. In contrast, at the eastern flank, the field is likely to be relatively more reversed compared with the magnetic field of the umbra and form X-point where magnetic reconnection could occur \citep{cheung2012}. This kind of asymmetry between two flanks of a light bridge is also consistent with fact that the chromospheric jet activities only cluster at one flank of the light bridge (see \figref{fig:pm}(d)). Small-scale flux emergence and jet activities are also reported in other granular light bridge examples, which support asymmetry of two flanks of a bridge \citep{louis2014,louis2015}.

In the speculated scenario, magnetic reconnections are triggered recurrently at the eastern flank, while repetitive compressions of magnetic field are driven by the same source at the western flank. This kind of magnetic morphology has been examined numerically, with repetitive reconnections triggered subsequently \citep{shibata1992, archontis2013}. These recurring small scale magnetic reconnections might be responsible for the waning of the bridge: magnetic ablation by reconnection could eventually lead to the disappearance of the bridge in a few days. \figref{fig:ts}(b) shows that the sub-minute oscillation power become weaker during this observational period. We will investigate this possible phenomenon in follow-up studies.

With regard to the periodicity at 0.5-1 minute, source evidence is inconclusive in current study. One possible source is the eigen mode of the magnetic resonator. Magnetic null point could host oscillatory reconnections due to plasma overshots. Oscillation time scale is estimated to be $\tau_\mathrm{osc}=2\ln S$ in unit of \alfven transit time from outer boundary into the diffusion region \citep{craig1991}, where Lundquist number $S$ is in the order of $10^6$ at chomospheric height. The \alfven transit time is estimated as $0.1L/V_\mathrm{A}$, where $L=100\unit{km}$ is the typical length scale of the reconnection site, where $0.1L$ is the length estimate of diffusion region \citep{craig1991}, and $V_\mathrm{A}=10\unit{km/s}$ is a typical \alfven speed at chromospheric height. The magnetic resonator will filter out the resonance period from the random perturbations \citep{mclaughlin2009,mclaughlin2012}. In our case, we estimate the oscillatory time scale at $\tau_\mathrm{osc}\sim30\unit{s}$ (see \citet{craig1991} for method), consistent with our measurement. 

In the proposed scenario, a dominant polarity is evaded by an emerging magnetic dipole. Such magnetic configuration have analogous models at both smaller and larger scales, such as the triggering mechanisms of X-ray jets \citep{archontis2013,sterling2015}, flares \citep{shibata2011} and mass ejections \citep{chen2011}. In our case, the reconnections is proposed to be triggered in a gentle and continuous manner; it also provide insight into the structure of the light bridge and their waning process (continuous ablation by repetitive magnetic reconnections)
 
This scenario also supports the nanoflare or microflare model. In the case of coronal loops, nanoflare heating \citep{parker1988} is proposed to occur rapidly at a spatial scale of 10 km or smaller, but it cannot be observed directly with current instruments. In the case of a sunspot contaminated by a light bridge, the interaction scale is up to the scale of 1000 km, and is within the resolution capability of available instruments. In this observation, a light bridge, normally a low atmosphere (low temperatures of 100,000 K or less) feature, is unusually observed at coronal channels (at about 2,000,000 K) for hours. Continuous heating is required to maintain the plasma at coronal temperatures. Moreover, we detected short-period (30-60 s) oscillations in the emission intensity only at hot coronal channels, with the timescale consistent with the high-frequency heating model \citep{klimchuk2015}. 

This study reports the first detection of two abnormal oscillation modes in a waning light bridge, our findings suggest a variety of new features. However, several of them cannot be conclusive owing to limited spatial and temporal resolutions. A coordinated observation with currently available instruments should be sufficient to provide better insights, e.g., New Solar Telescope (NST), Interface Region Imaging Spectrograph (IRIS). \textbf{We shall note that, although dynamic activities are prevailingly detected at two flanks of a light bridge, the spine usually looks  brighter in UV and EUV bandpasses. This might be related to the complex magnetic structure of the light bridge.}


\end{document}